\documentclass[useAMS,usenatbib]{mn2e}

\usepackage{graphicx}

\title[Relativistic Ejections in XTE J1752$-$223]{Transient Relativistic Ejections and Stationary Core in XTE~J1752$-$223}
\author[Yang et al.]{J.\,Yang$^{1}\thanks{E-mail:yang@jive.nl}$,
Z.\,Paragi$^{1,2}$,
S.\,Corbel$^{3}$,
L.I.\,Gurvits$^{1,4}$,
R.M.\,Campbell$^{1}$,
C.\,Brocksopp$^{5}$
\\
\\
$^{1}$Joint Institute for VLBI in Europe, Postbus~2,  7990~AA Dwingeloo, The Netherlands \\
$^{2}$MTA Research Group for Physical Geodesy and Geodynamics, POB~91, H-1521 Budapest, Hungary \\
$^{3}$Universit\'e Paris 7 Denis Diderot and Service d'Astrophysique, UMR AIM, CEA Saclay, F-91191 Gif-sur-Yvette, France \\
$^{4}$Department of Astrodynamics \& Space Missions, Delft University of Technology, Kluyverweg~1, 2629~HS Delft, The Netherlands  \\
$^{5}$Mullard Space Science Laboratory, University College London, Holmbury St Mary, Dorking, Surrey RH5~6NT, UK \\
}

\begin{document}

\date{Accepted 2011 August xx. Received 2011 July 20; in original form 2011 August xx}

\pagerange{\pageref{firstpage}--\pageref{lastpage}} \pubyear{2010}
\maketitle
\label{firstpage}

\begin{abstract}
The Galactic X-ray transient XTE~J1752$-$223 was shown to have properties of black hole binary candidates. As reported in our previous paper, we identified transient and decelerating ejecta in multi-epoch Very Long Baseline Interferometry (VLBI) observations with the European VLBI Network (EVN) and the NRAO Very Long Baseline Array (VLBA). Here we present new EVN and VLBA data in which a new transient ejection event and later a stationary component are identified. The latter is interpreted as a reappearance of the radio core/compact jet during the transition from soft to hard X-ray state. This component appears to be highly variable in brightness although effects of tropospheric instabilities might play a role too. We also re-analyze the earlier VLBI data and find that the transient ejecta closer to the core position has significantly higher proper motion, further strengthening the case for strongly decelerating ejecta on the scale of several hundred milli-arcsecond, never observed in X-ray binaries before. Although the distance of the source is not well constrained, it is clear that these ejectas are at least mildly relativistic at the early stages. Moreover, we show the large scale environment of the transient from the Westerbork synthesis array data recorded in parallel during the EVN run.

\end{abstract}


\section{Introduction}
\label{sec1}
The X-ray transient XTE~J1752$-$223 was discovered by the \emph{Rossi X-ray Timing Explorer} (\emph{RXTE}) on 2009 October 23. The 2010 outburst of XTE~J1752$-$223 was monitored by the \emph{RXTE} \citep{sha10}, the \emph{Monitor of All-sky X-ray Image} \citep[\emph{MAXI},][]{nak10}, and \emph{Swift} \citep{cur10} in the X-ray domain. The outburst exhibited an evolving behavior in agreement with what would be expected from a stellar mass black hole binary.

In the preceding low-hard X-ray state, the radio counterpart was identified by the ATCA (Australia Telescope Compact Array) with a flux density of $\sim$2~mJy and a flat spectrum consistent with that of a compact jet \citep{bro09}. XTE~J1752$-$223 started to transit toward a high-soft X-ray state around 2010 January 20 \citep[e.g.][]{sha10}. A 10-fold increase of radio flux density and a flat spectrum were observed by the ATCA on 2010 January 21 \citep{bro10}. The increased flux density suggested that a major ejection was imminent \citep{bro10} and, indeed, a transient jet was detected by the rapid-response EVN (European VLBI Newtork) observations on 2010 February 11. The transitions from a low-hard X-ray state toward a high-soft X-ray state are in fact, commonly associated with discrete ejection events \citep{cor04, fen04, fen09}. Together with follow-up VLBA (Very Long Baseline Array) observations in 2010 February, the transient jet was found to show significant deceleration, possibly due to the jet interaction with the surrounding medium \citep[][hereafter Paper~I]{yan10}.

\begin{figure*}
  \centering
  %Please increase the width if there is some space left in the final proof.
  \includegraphics[width=0.98\textwidth]{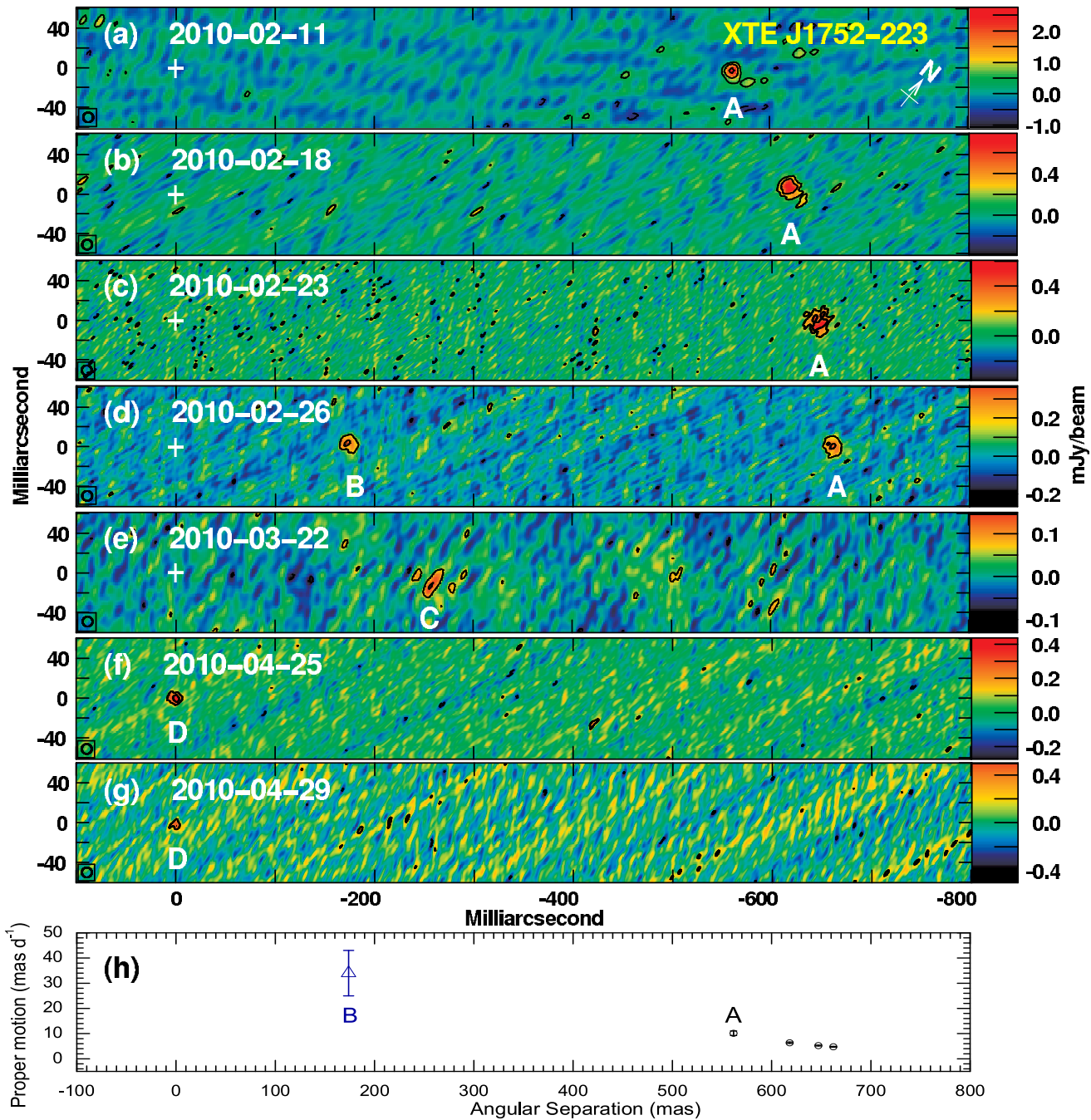} \\
  \caption{VLBI images of the transient jets components A, B, C, and the re-activated compact jet D in XTE~J1752$-$233. The bottom panel shows the proper motion versus angular separation for components A and B. All the images are centred at the compact jet (0,0) and rotated clockwise by 39$^\circ$. The linear scale is 3.5~au\,mas$^{-1}$ at the adopted distance of $3.5$~kpc. The contours start from 3$\sigma$ off-source noise level and increase by a factor of -1, 1, 2, 4. The size of the restoring beam is 10~mas.
  } \label{fig1}
\end{figure*}

In this paper we present results from the additional three epochs that show a new transient ejection event and later a stationary feature which we identify as the core. The observations and data reduction are briefly described in Section~\ref{sec2}. The Very Long Baseline Interferometry (VLBI) as well as the Westerbork Synthesis Radio Telescope (WSRT) imaging results are presented in Section~\ref{sec3}. The proper motion of a transient ejecta detected in one of our earlier observations is analyzed in Section~\ref{sec4}. We discuss our results on the compact jet region, the proper motion of the transient ejecta and their possible interaction with the large-scale medium in Section~\ref{sec5}.

\section{Observations and data reduction}
\label{sec2}
All the VLBI experiments observed at 5~GHz during the 2010 outburst of XTE~J1752$-$223 are listed in Table~\ref{tab1}. The observations before March 2010 were already presented in Paper~I. Additional EVN observations were carried out on 2010 March 22, including Effelsberg, the phased array WSRT, Onsala, Medicina, Noto, Toru\'n, and the Lovell telescope. There were two epochs of VLBA observations on 2010 April 25 and 29 using the whole telescope array. All the experiments used the same phase-referencing source (PMN~J1755$-$2232) and the same duty-cycle. The VLBI calibration procedure was identical as well, and it is described in Paper~I. We note that the new EVN observations show a weak extended jet in the calibrator source, but since the radio core accounts for $\sim$90\% of the emission, this extension did not have a significant effect on the phase calibration in the first four epochs. In the WSRT data, we did self-calibration on PMN~J1755$-$2232 and applied the solutions to the target.

\begin{table}
  \caption{The characteristics of VLBI images in Figure~\ref{fig1}.
}\label{tab1}
\scriptsize
\setlength{\tabcolsep}{4pt}
\centering
\begin{tabular}{ccccccc}
\hline
Map & Date       & Array
                         &$N_\mathrm{ant}$
                                & $t_\mathrm{obs}$
                                       &BW
                                                     & $S_\mathrm{peak}$                                                            \\
    & (yy-mm-dd) &       &    &  (h)   & (Mbps)  & (mJy~b$^{-1}$) \\
\hline
(a) & 10-02-11   & EVN   & 5  &  1.2   & 1024    & $2.67\pm0.26$   \\
(b) & 10-02-18   & VLBA  & 6  &  3.0   & 512     & $0.77\pm0.07$   \\
(c) & 10-02-23   & VLBA  & 7  &  6.0   & 512     & $0.60\pm0.07$   \\
(d) & 10-02-26   & VLBA  & 7  &  6.0   & 512     & $0.37\pm0.06$   \\
(e) & 10-03-22   & EVN   & 7  &  5.0   & 1024    & $0.14\pm0.03$   \\
(f) & 10-04-25   & VLBA  & 7  &  6.0   & 512     & $0.43\pm0.06$   \\
(g) & 10-04-29   & VLBA  & 8  &  1.6   & 512     & $0.50\pm0.10$   \\
\hline
\end{tabular}
\end{table}

\begin{table}
\caption{The results of the circular Gaussian model fitting for the VLBI-detected jet components in XTE~J1752$-$223. } \label{tab2}
\scriptsize
\centering
\begin{tabular}{ccrrc}
\hline
Comp.  & MJD      & Separation       & Position Angle     & Flux      \\
       & (d)      & (mas)~~~         &($^\circ$)~~~~~~~~  & (mJy)     \\
\hline
 A     & 55238.40 & $561.6\pm0.4$    &   $-51.3\pm0.1$    & 4.35      \\
 A     & 55245.58 & $618.2\pm0.5$    &   $-50.4\pm0.1$    & 2.20      \\
 A     & 55250.60 & $647.1\pm0.6$    &   $-51.2\pm0.1$    & 2.32      \\
 A     & 55253.60 & $662.2\pm0.8$    &   $-51.0\pm0.1$    & 1.05      \\
 B     & 55253.60 & $173.8\pm0.8$    &   $-49.9\pm0.3$    & 0.86      \\
 C     & 55277.23 & $259.6\pm1.8$    &   $-53.5\pm0.4$    & 0.75      \\
 D     & 55311.46 &   $0.0\pm0.6$    &                    & 0.43      \\
 D     & 55315.56 &   $3.2\pm1.2$    &    $-130\pm32$     & 0.50      \\
\hline
\end{tabular}
\end{table}

\section{VLBI and WSRT imaging results}
\label{sec3}
\subsection{VLBI detection of multiple jet features}
The VLBI total intensity images of XTE~J1752$-$223 are shown in Figure~\ref{fig1}. The associated image parameters and Gaussian model-fitting results are listed in Table~\ref{tab1} and \ref{tab2}. There are totally four jet features detected along the position angle of around $-51\degr$ and marked as A, B, C and D. Components A and B have been reported in Paper~I. The EVN image in Figure~\ref{fig1}e has achieved a sensitivity of $1\sigma=0.03$~mJy\,beam$^{-1}$, a factor of two better than other images, and revealed another weak ejecta component C with the peak brightness of 0.14~mJy\,beam$^{-1}$. Another jet feature, component D, is clearly seen on 2010 April 25 in Figure~\ref{fig1}f, but it was initially not detected on 2010 April 29. However, it is present clearly during part of the last observations, at exactly the same position as four days earlier. This indicates that the component has no detectable proper motion, and it is likely variable.

\begin{figure}
  \centering
  \includegraphics[width=0.48\textwidth]{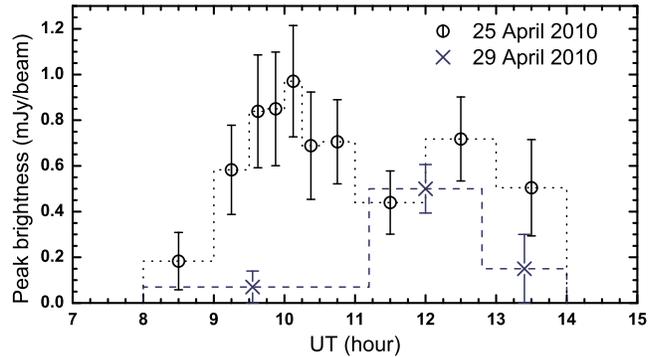} \\
  \caption{VLBI image peak brightness variation. The dotted and dashed lines are the length of each bin.
  } \label{fig2}
\end{figure}

The variation of the image peak brightness during the last two VLBI experiments is shown in Figure~\ref{fig2}. To pinpoint the peak time, a variable bin length was used. On 2010 April 25, the image peak brightness of XTE~J1752$-$223 boosted by a factor of five from the non-detection in the first 1-hour bin (8:00~--~9:00 UT) to 4$\sigma$ detection in a 15-minute bin (10:00~--~10:15 UT). After that, there was no significant decaying. A hint on variability was seen again on 2010 April 29. The source was not detected in the first 2.5 hours. While, it was detected at 5$\sigma$ during 11:10~--~12:48 UT as shown in Figure~\ref{fig1}g. The position difference from the previous detections is within the 3$\sigma$ error circle. Note that the variation of the $uv$-coverage does not affect the peak brightness in the case of a compact source.

\subsection{WSRT imaging of the XTE~J1752$-$223 field}

\begin{figure}
  \centering
  \includegraphics[width=0.48\textwidth, clip=true]{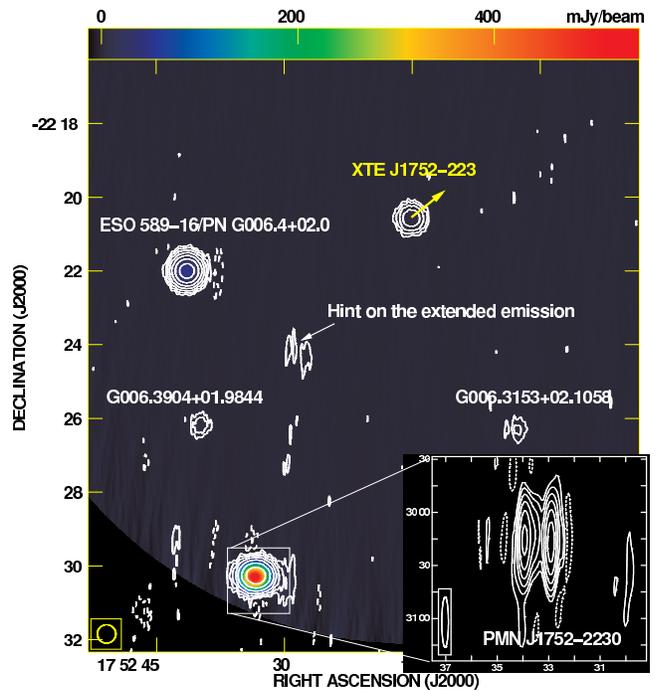} \\
  \caption{The WSRT total intensity image of the field of XTE~J1752$-$223. The yellow arrow highlights the jet direction. The contours start from $3\sigma=0.24$~mJy\,beam$^{-1}$ and increase by a factor of 2. The pseudo color image shows the prime beam-corrected intensity distribution.
  } \label{fig3}
\end{figure}

The WSRT image is shown in Figure~\ref{fig3}. To show possible extended emission, a circular restoring beam with a size 30\arcsec was used. The transient XTE~J1752$-$223 is a point source with a total flux density $2.8\pm0.2$~mJy. We also detected four other radio sources within the telescope primary beam. The small panel in Figure~\ref{fig3} shows the symmetric double morphology in the extragalactic source PMN~J1752$-$2230 \citep{sai04} obtained with the original synthesized beam: 28\arcsec\,$\times$\,4\arcsec. The eastern bright source is associated with the planetary nebula ESO~$589-16$ (PN~G006.4$+$02.0) at the distance of 5.1~kpc \citep{sta08}. The remaining sources are two stars associated with the \emph{Spitzer} GLIMPLSE \citep[Galactic Legacy Infrared Mid-Plane Survey Extraordinaire,][]{ben03} sources G006.3904$+$01.9844 and G006.3153$+$02.1058 within its $3\sigma=1\arcsec$ circle.

\section{Proper motion in transient jets}
\label{sec4}

Component A showed a proper motion decreased from $9.2\pm0.2$~mas\,d$^{-1}$ to $4.0\pm0.2$~mas\,d$^{-1}$, as revealed by Paper~I. To check if a significant proper motion can be detected in component B during the only experiment when it was detected, we split the VLBA data obtained on 2010 February 26 into various bins with different lengths. First we made two 3-hour bins and fitted for the position of the component. In addition, we probed bin sizes of 1 hour that provide larger separation between the centre of the first and the last bins. But in this case the four central bins had to be merged due to the low signal to noise ratio of the data (missing short spacings at that time range). Thus the position of component B was obtained for these additional three bins: the first hour, the central four hours, and the last 1 hour of data. We flagged out the high-noise data from Owens Valley (due to a warm receiver) and Brewster in the first half an hour (due to the low elevation); and left the radius, position angle and flux as free parameters to strengthen the conversion of the model fitting.

\begin{figure}
 \centering
  \includegraphics[width=0.23\textwidth, height=0.3\textwidth ]{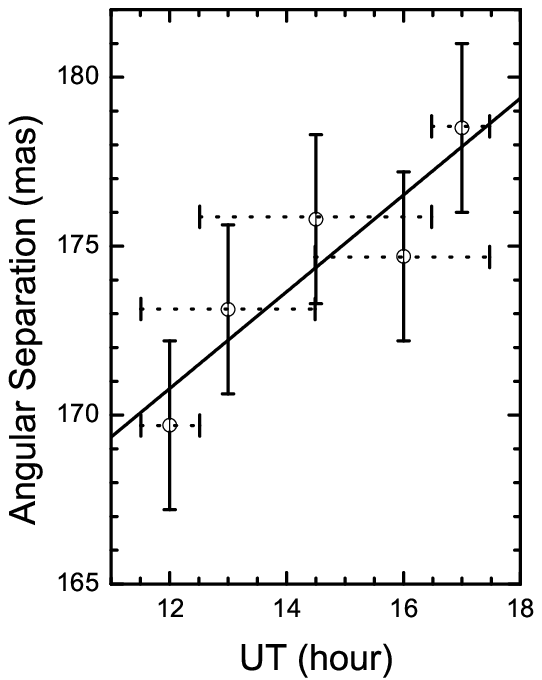}
  \includegraphics[width=0.23\textwidth, height=0.3\textwidth ]{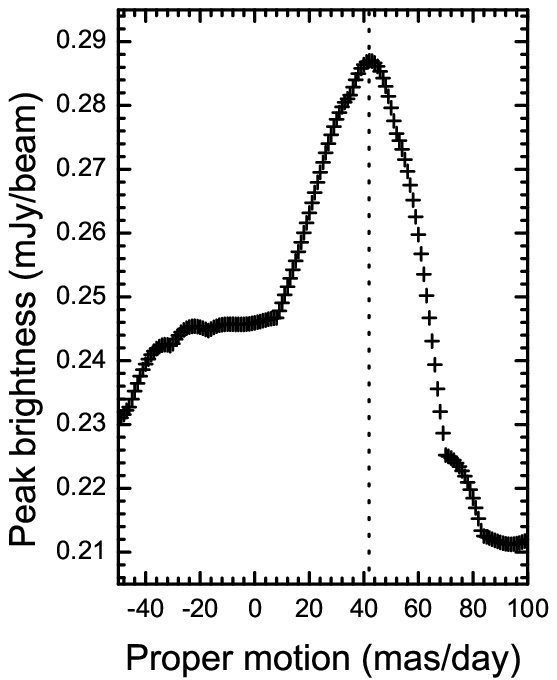} \\
 \caption{Proper motion in the XTE~J1752$-$223 component B. The straight line in the right panel shows the least-square fitting results: $34\pm9$~mas\,d$^{-1}$. Note that the horizontal dotted line crossing each data point gives the bin length. The right panel shows the variation of the peak brightness after applying a series of proper motion corrections along the jet direction, as described in the text.
  } \label{fig4}
\end{figure}

The angular separation versus time is displayed in the left panel of Figure~\ref{fig4} and listed in Table~\ref{tab3}. Component B had a position shift of $\sim$9~mas along jet direction within 5~hours. The linear least-square fit, plotted as a straight line, gives a proper motion of $34\pm9$~mas\,d$^{-1}$ with a reduced $\chi^2=0.4$. Since the precise position error is unknown, we scaled the errors with the square root of the reduce $\chi^2$ to estimate the proper motion error. Clearly, component B had a much higher proper motion than component A. The measurements for both components are plotted in the bottom panel of Figure~\ref{fig1} as well.

Alternatively, we searched for the proper motion by measuring the peak brightness of the image after the visibility data were corrected for a range of apparent velocities. Since the high proper motion obtained above smears the component structure during the observations lasting for 6 hours, one should expect to see an increased peak brightness near the real value. The correction along the jet direction was implemented via the AIPS \citep{gre03} task CLCOR\footnote{Note that the reference time is at 0:00~UT of the first observing day instead of the mid-point of the VLBI experiment in AIPS, and this had to be taken into account as well.}. We used pure natural weighting to make the dirty map, and looked for the peak position within component B. Before applying the proper motion correction to the $uv$-data, we removed component A with the AIPS task UVMOD from the calibrated visibility data, to avoid any potential interference from its sidelobes.

The variation of the image peak brightness versus the proper motion is plotted in the right panel of Figure~\ref{fig4}. The peak brightness has an increase of 17\% at a proper motion $42\pm7$~mas\,d$^{-1}$, marked with a dotted line. The proper motion error was estimated using its position error and the total observing time. The latter indirect measurement agrees well with the previous direct determination within 1$\sigma$ error.

Components C and B are unlikely to correspond to the same jet feature because the mean proper motion ($\sim$3.6~mas\,d$^{-1}$) between the two epochs is much smaller than that observed in component B. We also tried to estimate proper motion in component C using these two methods. Due to the limited sensitivity, we failed to achieve a meaningful result.

\begin{table}
\caption{The circular Gaussian model fitting results of component B in different UT ranges.} \label{tab3}
\scriptsize
\setlength{\tabcolsep}{8pt}
\centering
\begin{tabular}{cccrrrcr}
\hline
UT Range       & Separation       & Position Angle   & Flux      \\
(hh:mm)        & (mas)~~~         &($^\circ$)~~~~~~~~& (mJy)     \\
\hline
11:30~--~14:30 & $173.1\pm2.5$    &   $-50.1\pm0.9$  & 0.06      \\
14:30~--~17:30 & $174.7\pm2.5$    &   $-49.8\pm0.9$  & 0.10      \\
\hline
11:30~--~12:30 & $169.7\pm2.5$    &   $-50.0\pm0.9$  & 0.10      \\
12:30~--~16:30 & $175.8\pm2.5$    &   $-49.1\pm0.9$  & 0.07      \\
16:30~--~17:30 & $178.5\pm2.5$    &   $-49.4\pm0.9$  & 0.11      \\
\hline
\end{tabular}
\end{table}

\section{Discussion}
\label{sec5}
\subsection{The compact jet}
Component D was found to have no detectable proper motion with respect to the reference source at the last two VLBI epochs. The only explanation for this is that component D is the radio core of the system. We note that \citet{mil11} arrived to the same conclusion first: they independently analyzed data from the last two VLBA epochs, and showed that component D was positionally consistent with the optical counterpart of XTE~J1752$-$223, within a $1\sigma=46$~mas error circle. They had only a 3.5$\sigma$ detection on 2010 February 29. In our analysis, we obtained a considerably higher significance of the detection reaching 5$\sigma$, firmly establishing that component D is related to the core region/compact jet in XTE~J1752$-$223 and not related to a new transient ejecta.

Compared with the most sensitive EVN observations of 2010 March 22 during the X-ray soft state, the radio core re-brightened by a factor of $>$10. The non-detection in the earlier soft state is unlikely caused by the fast variability because the core was not detected on short time scales. The re-appearance provides a rare case of direct detection of the re-activation of radio emission in the compact jet during the soft to hard X-ray state transition, as described by the unified model for the coupling of accretion and ejection in black hole X-ray binaries \citep{fen09}. In fact, \emph{RXTE} observations on 2010 March 27 have revealed XTE~J1752$-$223 in transition from the soft to the hard state \citep{mun10}.

The compact jet appears highly variable. However, we note that the declination of the source is low and the strong apparent variation might be due to poor tropospheric conditions especially at the beginning and end of the observations, when the source elevation was very low. However the ATCA observations did show significant variations on timescales of days (Brocksopp et al., in prep.). In literature, strong daily variations were also observed in XTE~J1720$-$318 in the post low hard state \citep{bro05}. Significant flux density variation on shorter ($<1$~day) time scale were seen in GRS~1915$+$105 \citep{fen99} and XTE~J1650$-$500 \citep{cor04}.

\subsection{Relativistic ejections}
According to the new proper motion measurements, component B is a transient jet component moving toward the same direction as component A (cf. Paper~I), in agreement with the core location discussed above. Although we could not measure the proper motion of component C in the same way as component B, it lies between components A and B which indicates that there were at least three ejection events revealed by the VLBI observations during the 2010 outburst of XTE~J1752$-$223.

The source shows a profound one-sided morphology, unprecedented for X-ray black hole binaries with transient ejecta. We searched for the receding jet in the VLBI images including the deep EVN image. There were no receding ejecta found. If component A was ejected during the major X-ray outburst around MJD~55218, a mean proper motion of $\sim$28~mas\,d$^{-1}$ is required to explain the large angular separation from the assumed core. Considering the low proper motion of $\leq$10~mas\,d$^{-1}$ observed at the later stage, component A must have an extremely high proper motion at the early stage. The high proper motion observed in component B, an ejecta much closer to the core than component A, indirectly supports the association of component A with the major outburst. As a relativistic jet usually evolves with a decaying luminosity after the ejection, the non-detection of component B observed three days earlier is unlikely due to the insufficient sensitivity. Assuming component B was ejected around MJD~55250.6 (during another flare, Brocksopp et al. in prep.), its average proper motion is $\sim$58~mas\,d$^{-1}$, in agreement with our measurements.

Component B has a high apparent speed of $\sim$0.7~c if it is at 3.5~kpc as estimated by \citet{sha10} via the spectral-timing correlation scaling technique (model-dependent). However, if the source is located at a larger distance, e.g. near the Galactic centre, then XTE~J1752$-$223 might be another example of superluminal Galactic jets, such as GRS~1915$+$105 \citep{mir94, fen99} and GRO~J1655$-$40 \citep{tin95, hje95}. Taking into account that component B might have already somewhat decelerated by the time it was detected, XTE~J1752$-$223 is a promising Galactic superluminal source candidate. The resulting Doppler deboosting effect may be responsible for the non-detection of the receding transient jets, a rarely-seen property in stellar mass black hole binaries.

\subsection{A hint on the extended emission}
There might be an extended emission in rough agreement with the position angle of the transient jet between XTE~J1752$-$223 and G006.3904$+$01.9844 in Figure~\ref{fig3}. Such the extended emission might be a large-scale diffuse relic emission associated with the transient. As the limited sensitivity $3\sigma=0.25$~mJy~beam$^{-1}$, deep observations at lower frequencies are required to confirm its existence. A jet-blown bubble, $\sim$5~pc in diameter, is found in~Cygnus X-1 \citep{gal05}. It is inferred that there exists a cavity surrounding microquasars \citep{hao09}. As XTE~J1752$-$223 showed strong jet decelerations, it is a promising candidate for the deep observations to hunt for the extended emission region.

%\begin{table}
%\caption{ The variation of the VLBI image peak brightness with the observing time.} \label{tab3}
%%\scriptsize
%\setlength{\tabcolsep}{4pt}
%\centering
%\begin{tabular}{ccc}
%\hline
%Date        & UT Range       & Peak Brightness  \\
%(yyyy-mm-dd)&(hh:mm)         & (mJy/beam)       \\
%\hline
%2010-04-25  & 08:00~--~09:00 &	$0.18\pm0.14$  \\
%2010-04-25  & 09:00~--~09:30 &	$0.58\pm0.19$  \\
%2010-04-25  & 09:30~--~09:45 &	$0.84\pm0.25$  \\
%2010-04-25  & 09:45~--~10:00 &	$0.85\pm0.25$  \\
%2010-04-25  & 10:00~--~10:15 &	$0.97\pm0.24$  \\
%2010-04-25  & 10:15~--~10:30 & 	$0.69\pm0.23$  \\
%2010-04-25  & 10:30~--~11:00 &	$0.71\pm0.18$  \\
%2010-04-25  & 11:00~--~12:00 &	$0.44\pm0.14$  \\
%2010-04-25  & 12:00~--~13:00 &	$0.72\pm0.18$  \\
%2010-04-25  & 13:00~--~14:00 &	$0.50\pm0.21$  \\
%\hline
%2010-04-29  & 08:00~-~11:08 & 	$0.07\pm0.07$  \\
%2010-04-29  & 11:10~-~12:48 & 	$0.50\pm0.11$  \\
%2010-04-29  & 12:48~-~14:00 &   $0.15\pm0.15$  \\
%\hline
%\end{tabular}
%\\
%\end{table}

\section[]{Conclusions}
We investigated the proper motion of the closest ejecta to the assumed core among the detected transient jet components in the 2010 outburst of XTE~J1752$-$223. Besides measuring proper motion through splitting the experiment into a few time bins, we looked for a proper motion via monitoring the variation of the image peak brightness of the proper motion-corrected visibility data. Both ways give a very high proper motion and imply that the ejections are at least mildly relativistic. We also reported the detection of another new ejecta observed at the end of 2010 March. We provided another independent VLBI identification of the compact jet and have shown that it might be highly variable. Moreover, we found a hint on the large-scale extended emission in the field of XTE~J1752$-$223 with the WSRT observations.

\section*{Acknowledgments}
\footnotesize
We thank the EVN Programme Committee and the VLBA Proposal Selection Committee for prompt approval of our observations. e-VLBI developments in Europe were supported by the EC DG-INFSO funded Communication Network Developments project EXPReS. The National Radio Astronomy Observatory is a facility of the National Science Foundation operated under cooperative agreement by Associated Universities, Inc. The EVN is a joint facility of European, Chinese, South African and other radio astronomy institutes funded by their national research councils. The WSRT is operated by ASTRON with support from the Netherlands Foundation for Scientific Research.

\bsp
\label{lastpage}

\normalsize

\begin{thebibliography}{99}
\bibitem[\protect\citeauthoryear{Benjamin et al.}{2003}]{ben03} Benjamin~R.A. et al., 2003, PASP, 115, 953

\bibitem[\protect\citeauthoryear{Brocksopp et al.}{2005}]{bro05} Brocksopp~C. et al., 2005, MNRAS, 356, 135

\bibitem[\protect\citeauthoryear{Brocksopp et al.}{2009}]{bro09} Brocksopp~C. et al., 2009, Astronomer's Telegram, 2278

\bibitem[\protect\citeauthoryear{Brocksopp et al.}{2010}]{bro10} Brocksopp~C. et al.,  2010, Astronomer's Telegram, 2400

\bibitem[\protect\citeauthoryear{Corbel et al.}{2004}]{cor04} Corbel~S. et al., 2004, ApJ, 617, 1272

\bibitem[\protect\citeauthoryear{Curran et al.}{2010}]{cur10} Curran~P.A. et al., 2011, MNRAS, 410, 541

\bibitem[\protect\citeauthoryear{Fender et al.}{1999}]{fen99} Fender~R.P. et al., 1999, MNRAS, 304, 865

\bibitem[\protect\citeauthoryear{Fender et al.}{2004}]{fen04} Fender~R.P. et al., 2004, MNRAS, 355, 1105

\bibitem[\protect\citeauthoryear{Fender et al.}{2009}]{fen09} Fender~R.P. et al., 2009, MNRAS, 396, 1370

\bibitem[\protect\citeauthoryear{Gallo et al.}{2005}]{gal05} Gallo~E. et al., 2005, Nature, 436, 819

\bibitem[\protect\citeauthoryear{Greisen}{2003}]{gre03} Greisen~E.W., 2003, in Heck~A., ed., Information Handling in Astronomy: Historical Vistas, ASSL, 285, 109

\bibitem[\protect\citeauthoryear{Hao \& Zhang}{2009}]{hao09} Hao~J.F. \& Zhang~S.N., 2009, MNRAS, 702, 1648

\bibitem[\protect\citeauthoryear{Hjellming \& Rupen}{1995}]{hje95} Hjellming~R.M., Rupen~M.P., 1995, Nat, 375, 464

\bibitem[\protect\citeauthoryear{Miller-Jones et al.}{2011}]{mil11} Miller-Jones~J.C.A. et al., 2011, MNRAS, 415, 306

\bibitem[\protect\citeauthoryear{Mirabel \& Rodr\'iguez}{1994}]{mir94} Mirabel~I.F., Rodr\'iguez~L.F., 1994, Nat, 371, 46

\bibitem[\protect\citeauthoryear{Mu\~noz Darias et al.}{2010}]{mun10} Mu\~noz~Darias~T. et al., 2010, Astronomer's Telegram, 2518

\bibitem[\protect\citeauthoryear{Nakahira et al.}{2010}]{nak10} Nakahira~S. et al., 2010, PASJ, 62, L27

\bibitem[\protect\citeauthoryear{Saikia et al.}{2004}]{sai04} Saikia~D.J. et al., 2004, MNRAS, 354, 827

\bibitem[\protect\citeauthoryear{Shaposhnikov et al.}{2010}]{sha10} Shaposhnikov~N. et al., 2010, ApJ, 723, 1817

\bibitem[\protect\citeauthoryear{Shepherd et al.}{1994}]{she94} Shepherd~M.C. et al., 1994, BAAS, 26, 987

\bibitem[\protect\citeauthoryear{Stanghellini et al.}{2008}]{sta08} Stanghellini~L. et al. 2008, ApJ, 689, 194

\bibitem[\protect\citeauthoryear{Tingay et al.}{1995}]{tin95} Tingay~S.J. et al., 1995, Nat, 374, 141

\bibitem[\protect\citeauthoryear{Yang et al.}{2010}]{yan10} Yang~J. et al., 2010, MNRAS, 409, L64

\end{thebibliography}
\end{document}